\documentstyle[fleqn]{article}
\title          {Quantum dynamics of $N=1$, $D=4$ supergravity compensator}
\author{I.L.Buchbinder and A.Yu.Petrov\\
        Department of Theoretical Physics\\
        State Pedagogical Institute\\
        Tomsk 634041, Russia}
\date{}
\begin {document}
\maketitle
\begin{flushright}
                                                       TSPI-TH32/96
\end{flushright}

{\it Talk given by I.L.Buchbinder. To be published in the Proceedings of the
SUSY96 conference}

\newpage

       {\Large  Quantum dynamics of $N=1$, $D=4$ supergravity compensator} \\

\begin{center}
        I.L.Buchbinder and A.Yu.Petrov \\
        Department of Theoretical Physics\\
        Tomsk State Pedagogical Institute\\
        Tomsk 634041, Russia\\
\end{center}

\begin{flushright}
                                                       TSPI-TH32/96
\end{flushright}

\begin{abstract}
 A new $N=1$ superfield theory in $D=4$ flat superspace is suggested.
 It describes dynamics of supergravity compensator and can be considered
 as a low-energy limit for $N=1$, $D=4$ superfield supergravity. The theory
 is shown to be renormalizable in infrared limit and infrared free. A
 quantum effective action  is investigated  in infrared domain.
\end{abstract}

It is well known that the superfield supergravity has been constructed in refs.
[1-2] as a dynamical theory of vector superfield and chiral and antichiral
superfield compensators in $N=1$,
$D=4$ curved superspace. A general consideration of quantum aspects of
superfield supergravity has been carried out in refs. [3].

Purpose of this paper is to develop a simplified theory describing dynamics of
only compensator superfields in flat superspace.
We show that such a theory can be treated as a natural low-energy limit of
quantum supergravity with matter and possesses remarkable properties in
infrared domain.

The model under consideration is based on an idea of gravity induced by
conformal anomaly of matter fields in curved space-time [4] (see also [5]).
A superconformal anomaly of matter superfields in $N=1$ curved superspace has
been found in ref. [6] and
a superfield action generating this anomaly has been constructed in ref.[7]
(see also [8]).
We investigate a model an action of which is a sum
of the anomaly generating action and the action of $N=1, D=4$ superfield
supergravity (see f.e. [8]). Being transformed to conformally flat superspace
the action of the model takes the form
\begin {eqnarray}
&S&=\int d^8z
    (-\frac{Q^2}{2{(4\pi)}^2}\bar{\sigma}\Box\sigma+\bar{D}^{\dot\alpha}\bar
{\sigma}D^{\alpha}\sigma\times\nonumber\\
&\times&({\xi_1\partial_{\alpha}}_{\dot{\alpha}}(\sigma+\bar{\sigma})
+\xi_2\bar{D}_{\dot{\alpha}}\bar{\sigma} D_{\alpha}\sigma)-\nonumber\\
&-&\frac{m^2}{2} e^{\sigma+\bar{\sigma}})+(\Lambda\int d^6z e^{3\sigma}+h.c.)
\end {eqnarray}
Here $D_{\alpha}$, $\bar{D}_{\dot{\alpha}}$, $\partial_{\alpha\dot{\alpha}}$
are the flat supercovariant derivatives,
$\sigma=\ln\Phi$ and $\Phi$ is a chiral compensator of $N=1$, $D=4$ supergravity.
$Q^2$, $\xi_1$, $\xi_2$, $m^2$, $\Lambda$ are the arbitrary parameters of the
model playing a role of the couplings.

We begin with consideration of renormalization structure of the model (1).
One can prove that the superficial degree of divergences $\omega$ has the form
[10]
\begin {equation}
\omega\leq 2-3L_c-2V_3-3V_4
\end {equation}
Here $V_3$ is a number of vertices proportional to $m^2$,
$V_4$ is a number of vertices proportional to $\Lambda$, $L_c$ is a number of
lines corresponding to the propagators $<\sigma\sigma>$ and
$<\bar{\sigma}\bar{\sigma}>$.

The condition of divergence $\omega\geq 0$ means that the divergent
supergraphs cannot contain the vertices proportional to $\Lambda$ and the above
lines. They can include no more than one vertex proportional to $m^2$.

As a result we have some sort of non-renormalization
theorem according to which the vertex proportional to $\Lambda$ is
always finite.

Let us consider now an one-loop renormalization. The supergraphs leading to
one-loop divergences are given by the Fig.1, Fig.2 and Fig.3

\begin{picture}(100,100)
\put(50,50){\circle{40}}
\put(30,50){\line(-1,-1){20}}
\put(15,40){\line(0,-1){10}}
\put(10,20){$D^{\alpha}$}
\put(35,70){\line(0,-1){10}}
\put(30,70){$\bar{D}_{\dot{\alpha}}$}
\put(30,50){\line(-1,1){20}}
\put(15,70){\line(0,-1){10}}
\put(35,40){\line(0,-1){10}}
\put(30,20){$D_{\alpha}$}
\put(10,75){$\bar{D}^{\dot{\alpha}}$}
\put(70,50){\line(1,0){20}}
\put(80,40){$\partial_{\beta\dot{\beta}}$}
\put(80,55){\line(0,-1){10}}
\put(65,40){\line(0,-1){10}}
\put(65,70){\line(0,-1){10}}
\put(70,75){$D^{\beta}$}
\put(70,20){$\bar{D}^{\dot{\beta}}$}
\put(45,80){$G_{+-}$}
\put(45,20){$G_{-+}$}
\put(40,0){Fig.1}
\end{picture}
\begin{picture}(100,100)
\put(50,50){\circle{40}}
\put(30,50){\line(-1,-1){20}}
\put(30,50){\line(-1,1){20}}
\put(70,50){\line(1,-1){20}}
\put(70,50){\line(1,1){20}}
\put(15,40){\line(0,-1){10}}
\put(10,20){$D^{\alpha}$}
\put(15,70){\line(0,-1){10}}
\put(10,75){$\bar{D}^{\dot{\alpha}}$}
\put(85,70){\line(0,-1){10}}
\put(85,40){\line(0,-1){10}}
\put(85,20){$D^{\beta}$}
\put(85,75){$\bar{D}^{\dot{\beta}}$}
\put(35,40){\line(0,-1){10}}
\put(30,20){$D_{\alpha}$}
\put(35,70){\line(0,-1){10}}
\put(30,70){$\bar{D}_{\dot{\alpha}}$}
\put(65,40){\line(0,-1){10}}
\put(65,70){\line(0,-1){10}}
\put(70,75){$D_{\beta}$}
\put(70,20){$\bar{D}_{\dot{\beta}}$}
\put(45,80){$G_{+-}$}
\put(45,20){$G_{-+}$}
\put(40,0){Fig.2}
\end{picture}

\begin{center}
\begin{picture}(100,100)
\put(50,50){\circle{40}}
\put(30,50){\line(-1,0){20}}
\put(70,50){\line(1,0){20}}
\put(15,55){\line(0,-1){10}}
\put(80,55){\line(0,-1){10}}
\put(80,40){$\partial_{\alpha\dot{\alpha}}$}
\put(10,60){$\partial_{\beta\dot{\beta}}$}
\put(65,40){\line(0,-1){10}}
\put(65,70){\line(0,-1){10}}
\put(70,75){$D^{\alpha}$}
\put(70,20){$\bar{D}^{\dot{\alpha}}$}
\put(35,40){\line(0,-1){10}}
\put(30,20){$D^{\beta}$}
\put(35,70){\line(0,-1){10}}
\put(30,70){$\bar{D}^{\dot{\beta}}$}
\put(45,80){$G_{+-}$}
\put(45,20){$G_{-+}$}
\put(40,0){Fig.3}
\end{picture}
\end{center}
Here $G_{+-}$ and $G_{-+}$ are the $<\sigma\bar{\sigma}>$- and\\
$<\bar{\sigma}\sigma>$ propagators respectively. After straightforward
calculations of above supergraphs in framework of dimensional reduction
regularization we obtain
\begin{eqnarray}
Q^2_{(0)}&=&\mu^{-\epsilon} Z_Q Q^2,\ \ \ \xi_{(0) 1,2}= \mu^{-\epsilon}
Z \xi_{1,2}\\
Z_Q&=&1+\frac{2^{14} 3^2 \pi^6 \xi_1^2}{Q^6\epsilon},\ \
Z=1+\frac{2^{11} 3^2\pi^2\xi_2}{Q^4\epsilon}\nonumber
\end{eqnarray}
The eqs. (3) lead to the following equations for running couplings
\begin{eqnarray}
\frac{d\xi_1(t)}{dt}&=&a\frac{\xi_1(t)\xi_2(t)}{Q^4(t)};\
\frac{d\xi_2(t)}{dt}=a\frac{\xi_2^2(t)}{Q^4(t)}\nonumber\\
\frac{d Q^2(t)}{dt}&=&b\frac{\xi_1^2(t)}{Q^4(t)}
\end{eqnarray}
where $a=2^{11}3^2\pi^2$, $b=3^2 2^{14}\pi^4$.
These equation show that the $\xi_1(t)= \xi_2(t)=0$
is an infrared fixed point and $Q^2(t)\to Q^2=const$
in infrared domain $t\rightarrow-\infty$. 

Now we consider a renormalization of the coupling $m^2$ in infrared limit where
$\xi_1=\xi_2=0$. 
The corresponding divergent supergraphs are given by Fig.4.

\begin{picture}(100,100)
\put(50,50){\circle{40}}
\put(30,50){\line(-1,-1){20}}
\put(30,50){\line(-1,1){20}}
\put(20,45){$\vdots$}
\put(75,50){$G_{+-}$}
\put(40,0){Fig.4}
\end{picture}\\
Here the dots mean external $\sigma$, $\bar{\sigma}$-lines.
A straightforward calculation lead to
\begin{equation}
m^2_0=Z_{m^2} m^2;\ \ Z_{m^2}=1+\frac{2}{Q^2\epsilon}
\end{equation}

The equation for running coupling $m^2(t)$ are written in the form
\begin{eqnarray}
\frac{d m^2(t)}{dt}&=&(2-2\alpha+\frac{2\alpha^2}{Q^2})m^2(t)
\end{eqnarray}
As a result the running coupling $m^2(t)$ vanishes in infrared limit. 

Taking into account everything above we see that the model with the action
\begin{eqnarray}
\label{ir}
&S&=\int d^8 z(-\frac{1}{2}\frac{Q^2}{16\pi^2}\bar{\sigma}\Box\sigma)+\\
&+&(\frac{\Lambda}{\alpha^2}\int d^6 z e^{3\alpha\sigma}+ h.c.)\nonumber
\end{eqnarray}
corresponds to finite in infrared domain theory which can be treated
as a natural infrared limit of the quantum supergravity with matter. 

We investigate now a structure of effective action for the model (\ref{ir}).
Since the model is finite we face a problem of effective action in finite 
quantum field theory.

An effective action in $N=1$, $D=4$ superfield theory is written in the form
\begin{equation}
\Gamma=\int d^8 z L+(\int d^6 z {\cal L}_c +h.c.)
\end{equation}
Here $L$ is called the general effective lagrangian,
${\cal L}_c$ is called the chiral effective lagrangian.
The $L$ depends on the field $\sigma$ and $\bar{\sigma}$ and their
supercovariant derivatives.
In particular there can be a term independent of the derivatives. 
This term is called a kahlerian effective potential $K$ (see the discussion in
[8,9]).
The ${\cal L}_c$ depends only on $\sigma$ and its space-time derivatives of 
any order.
We represent $L(\sigma,\bar{\sigma})$ and ${\cal L}_c(\sigma)$ 
in the form of power expansion in derivatives of $\sigma$, $\bar{\sigma}$ and 
will find the lower contributions in this expansion.

The action (7) is
invariant under transformations
\begin{equation}
\sigma\to \sigma+\gamma, \bar{\sigma}\to \bar{\sigma}+\beta,
\Lambda\to e^{-3\gamma}\Lambda, \bar{\Lambda}\to e^{-3\beta}\Lambda
\end{equation}
One can expect that the effective action should be also invariant under these 
transformations (see discussion in ref.[10]). Taking into account the dimensions
of $\sigma$, $\Lambda$, $\bar{\Lambda}$, ${\cal L}$ and ${\cal L}_c$ we obtain
the lower contributions to ${\cal L}$ and ${\cal L}_c$ in the form
\begin{eqnarray}
K= c{\Big(\frac{\Lambda}{Q^2}\Big)}^{2/3}
e^{\sigma+\bar{\sigma}}\nonumber\\
{\cal L}_c=\frac{\Lambda^{1/3}}{Q^{2/3}}
\{( c_1 e^{-\sigma} +c_2 e^{2\sigma}+\\
+c_3 e^{-\sigma}+c_4 e^{-4\sigma})
\partial^m\sigma \partial_m\sigma+\nonumber\\
+\Box\sigma (c_3 e^{-\sigma} +c_4 e^{-4\sigma})\}\nonumber
\end{eqnarray}
where $\bar{\Lambda}=\Lambda$; $c$, $c_1$, $c_2$, $c_3$, $c_4$ are arbitrary 
dimensionless constants.
Explicit form of these constants can be found by straightforward calculations in
one-loop approximation (see ref. [10], where a special technique for calculation 
of one-loop effective action for the model (7) has been developed).

As a result the effective action with lower contributions has a form of classical
action (7) and quantum corrections defined by $K$ and ${\cal L}_c$. One can prove
that for the superfields $\sigma$, $\bar{\sigma}$ slowly varying in space-time 
this effective action can be rewritten in the form
\begin{equation}
\Gamma=\int d^8 z \phi\bar{\phi}+\big(\lambda \int d^6 z {\phi}^3+ h.c.)
\end{equation}
where $\lambda=Q^2 c^{-3/2}$,
$\phi={(\frac{\Lambda}{Q^2})}^{1/3} e^{\sigma}$.
It is evident that the action (11) coincides with the standard action of the 
massless Wess-Zumino model. 

Let us summarize our results.
We have formulated a new model of chiral superfield in $N=1$, $D=4$ flat
superspace. This model is generated by superconformal anomaly of matter
superfields and can be considered as a simplified model of quantum supergravity
in low-energy domain. 

We have evaluated the superficial degree of divergences, calculated the one-loop superfield
counterterms, investigated the equations for running couplings and shown that
the model is infrared free and moreover it is finite in infrared limit. 
As a result we have constructed a supersymmetric formulation of the model by
Antoniadis and Mottola [4].

An interesting feature of the model is a non-renormalization theorem according
to which the vertex $\Lambda\int d^6 z e^{3\sigma}$ has no divergent corrections.
Since the superfield $\sigma$ is not renormalized in this model (instead of this
the parameter $Q^2$ is renormalized) we obtain that the
parameter $\Lambda$ (cosmological constant) is always finite. 

We have considered the effective action in our model in infrared domain
and calculated the lower contributions to 
kahlerian effective potential and chiral effective lagrangian.

Details of calculation connected with renormalization structure, one-loop 
counterterms, equations for running couplings and effective lagrangians are 
given in refs. [10].

\underline{Acknowledgements.}
 The work was supported by Russian Foundation for Basic Research under the
project No.96-02-16017.


\begin{thebibliography}{10}
\bibitem{OS}
  V. I. Ogievetsky, E. S. Sokatchev. Phys. Lett. B79, 222 (1978).
  Yad. Fiz. (Sov. J. Nucl. Phys) 31, 264
  (1980);32, 862 (1980); 32, 870(1980); 32, 1142 (1980).
\bibitem{SG}
  W. Siegel, S. J. Gates. Nucl. Phys. B147, 77 (1979).
\bibitem{GS}
  M. T. Grisaru, W.Siegel. Nucl. Phys. B187, 149 (1981); B201, 292 (1982).
\bibitem{AM}
  A. M. Polyakov. Phys.Lett. B103, 207 (1981).
  R.J.Reigert. Phys.Lett. B134, 56 (1984)
  E.S.Fradkin, A.A.Tseytlin. Phys.Lett. B134, 187 (1984).
  I. Antoniadis, E. Mottola. Phys. Rev. D45, 2013 (1992).
\bibitem{BO}
  I. L. Buchbinder, S.D.Odintsov, I. L. Shapiro. Effective Action in Quantum 
Gravity, IOP Publishing Ltd, Bristol and Philadelphia, 1992.
\bibitem{BKY}
  I.L.Buchbinder and S.M.Kuzenko. Nucl.Phys. B274, 653 (1986)
\bibitem{BK}
  I. L. Buchbinder and S. M. Kuzenko. Phys. Lett. B202, 233 (1988).
\bibitem{BK0}
  I. L. Buchbinder, S. M. Kuzenko. Ideas and Methods of Supersymmetry and
  Supergravity, IOP Publishing Ltd, Bristol and Philadelphia, 1995.
\bibitem{BK1}
  I. L. Buchbinder, S. M. Kuzenko, J. V. Yarevskaya. Nucl. Phys. B411, 665
  (1994).
  I. L. Buchbinder, S. M. Kuzenko, A. Yu. Petrov. Phys. Lett. B321, 372 (1994).
\bibitem{hp1}
  I. L. Buchbinder, A. Yu. Petrov. 
Quantum dynamics of N=1, D=4 supergravity chiral compensator. hep-th 9501125
(Class. Quant. Grav., to be published);
  I. L. Buchbinder, A. Yu. Petrov. 
On structure of effective action in four-dimensional quantum dilaton 
supergravity. hep-th 9604154.
\end{thebibliography}
\end{document}